\documentclass[sigconf]{acmart}
\usepackage{subcaption}
\usepackage{mathtools}
\usepackage{multirow}
\usepackage{hhline} 
\renewcommand\footnotetextcopyrightpermission[1]{} 
\AtBeginDocument{%
  \providecommand\BibTeX{{%
    \normalfont B\kern-0.5em{\scshape i\kern-0.25em b}\kern-0.8em\TeX}}}

\begin{document}

\title{Modeling Document Interactions for Learning to Rank with Regularized Self-Attention}

\author{Shuo Sun}
\affiliation{%
  \institution{Johns Hopkins University}
    }
\email{ssun32@jhu.edu}

\author{Kevin Duh}
\affiliation{%
  \institution{Johns Hopkins University}
}
\email{kevinduh@cs.jhu.edu}

\begin{abstract}
Learning to rank is an important task that has been successfully deployed in many real-world information retrieval systems.
Most existing methods compute relevance judgments of documents independently, without holistically considering the entire set of competing documents. 
In this paper, we explore modeling documents interactions with self-attention based neural networks.
Although self-attention networks have achieved state-of-the-art results in many NLP tasks, we find empirically that self-attention provides little benefit over baseline neural learning to rank architecture.
To improve the learning of self-attention weights, We propose simple yet effective regularization terms designed to model interactions between documents.
Evaluations on publicly available Learning to Rank (LETOR) datasets show that training self-attention network with our proposed regularization terms can significantly outperform existing learning to rank methods.
\end{abstract}

\maketitle

\date{}

\section{Introduction}
Learning to rank has attracted much attention in the research community, where the focus has been on developing loss objectives that effectively optimizes information retrieval (IR) metrics. 
The general idea is to fit a global relevance function $f(\cdot)$ on a training set that consists of queries, sets of documents, and their desired rankings. 
In particular, given a query $q$, a set of documents $\{d_1, d_2, \dots, d_n\}$, existing methods learn a function $f(q,d_i)$ that gives higher scores to documents that rank better. 
There are three common approaches:
\begin{enumerate}
    \item \textbf{Pointwise approaches} optimize the relevance score of a single document without considering other documents in the same set.
    \item \textbf{Pairwise approaches} optimize the ranking between pairs of documents, such that $f(q,d_i) > f(q,d_j)$ if $d_i$ ranks better than $d_j$.
    \item \textbf{Listwise approaches} directly attempts to optimize the target IR metric (such as MAP or NDCG), which are based on the entire set of document scores.
\end{enumerate}

Importantly, all these approaches focus on the loss objective during the \emph{training phase}. Whether the objective is pairwise or listwise, the function $f(q,d_i)$ computes relevance scores for each document $d_i$ independently at \emph{test inference time}. 

We propose to formulate the relevance function based on the set of documents to be ranked: $f(q,d_i, \{d_1, d_2, \dots, d_n\})$.\footnote{The notation will be described more precisely later. For now, the point is to illustrate the difference between modeling $f(q,d_i)$ independently for each $d_i$, versus adding the full document set in $f(q,d_i,\{d_1, d_2, \dots, d_n\})$. Suppose some competing documents are dropped, $f(q,d_i)$ will output the same relevance score, whereas $f(q,d_i,\{d_1, d_2, \dots, d_n\})$ will automatically adapt.}
This is more similar to how humans might rank documents at test time: multiple competing documents are reviewed before assigning the relevance score to $d_i$.

Recently, self-attention has been successfully applied to many tasks such as machine translation \cite{vaswani2017attention} and natural language inference \cite{devlin2018bert, liu2019roberta, lan2019albert}. 
As self-attention has the capability to establish direct connections among elements within a set, it is a suitable mechanism that can model interactions among documents.
The use of self-attention on set of documents allows the model to adjust scores based on other competing documents. 

However, experiment results on benchmark datasets show that ListNet with self-attention only performs marginally better.
Deeper analyses of attention weights reveal that self-attention alone is not effective at modeling document interactions.
In section 2, we propose regularization terms that can push the model towards learning meaning weights that can better model interactions between documents.

We evaluated our model on the popular Yahoo, MSLR-WEB and Istella LETOR datasets and show that neural networks with properly regularized self-attention weights can significantly outperform existing strong ensemble tree and neural network baselines.

\section{Model Description}

Given a query q, a set of documents $\{d_1, d_2, \dots, d_n\}$ and a feature extraction function, $\phi$, the input to a learning to rank model is a set of feature vectors:
\begin{equation}
S = \{\phi(q, d_1), \phi(q, d_2), \dots, \phi(q, d_n)\}
\end{equation}
We want to model a ranking function $f$ such that:
\begin{equation}
f(S) \rightarrow [s_1,s_2, \dots, s_n]
\end{equation}
where $s_i$ is the predicted relevance score for document $d_i$.
In this notation, $f(S)$ now has a vector output of dimension $n$, where each element represents the relevance score for a document. 

Ideally, we want $s_i$ to be sorted in the same order as the desired ranking. At test inference time, we compute all relevance scores then sort the documents according to these scores. 

\subsection{ListNet}
Our starting point for modeling $f$ is the ListNet\cite{cao2007learning} algorithm. ListNet is a strong neural learning to rank algorithm which optimizes a listwise objective function. 
Due to the combinatorial nature of the ranking tasks, popular metrics such as NDCG \cite{jarvelin2002cumulated} and ERR \cite{chapelle2009expected} are not differentiable with respect to model parameters and consequently gradient descent based learning algorithm cannot be directly used for optimization. 
Therefore, ListNet optimizes a surrogate loss function which is defined below: 

 Given predicted relevance judgments $f(S) = \{s_1,s_2,\dots, s_n\}$ and ground truth $R(S) = \{r_1,r_2,\dots, r_n\}$. The top one probability of document $d_j$ based on f is:

\begin{equation}
  P_f(d_j) = \dfrac{e^{s_j}}{\sum_k e^{s_k}}
\end{equation}
and the top one probability of document $d_j$ based on R is:
\begin{equation}
  P_R(d_j) = \dfrac{e^{r_j}}{\sum_k e^{r_k}}
\end{equation}
Loss is defined as the cross entropy between the top one probability distribution of predicted scores and the top one probability distribution of ground truth:
\begin{equation}
  L = -\sum_{i=1}^nP_R(d_i)log(P_f(d_i))
\end{equation}

\subsection{Self-Attention (SA)}
Self-attention is an attention mechanism which learns to represent every element in a set by a weighted sum of every other elements within the same set. 
Self-attention based neural networks have found success in many NLP tasks such as machine reading, machine translation and sentence representations learning \cite{lin2017structured, vaswani2017attention, cheng2016long, devlin2018bert}. 

The input to the self-attention layer is a set of vector representations:
\begin{equation}
  V = [v_1, v_2, \dots, v_n]
\end{equation}
here $v_i$ is a d-dimensional vector representation of the i-th document, $v_i \in \mathbb{R}^d$. V is a $n \times d$ matrix, which concatenates the vector representations of the n documents. The output of the self-attention layer is:
 \begin{equation}
    V' = \sigma\left((VW_q)(VW_k)^\intercal\right)(VW_v)\label{eq:1}
 \end{equation}
 where $D'_i \in \mathbb{R}^{n\times h}$, $D \in \mathbb{R}^{n\times d}$, $W^i_q, W^i_k, W^i_v \in \mathbb{R}^{d\times h}$ and $\sigma$ is the sigmoid function.
 $W^i_q, W^i_k$ and $W^i_v$ are trainable weight matrices.

\subsection{ListNet + Self-Attention (SA)}
\begin{figure*}[htbp!]
    \centering
    \includegraphics[width=0.8\linewidth]{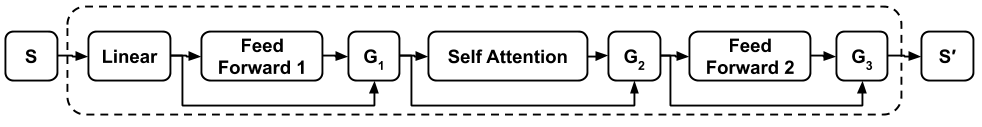}
    \captionsetup{width=0.8\linewidth}
    \caption{A document encoder consisting of two feed forward layers and a self-attention layer. $G_1$, $G_2$, $G_3$ are highway connections \cite{srivastava2015highway}.}
    \label{fig:docencoder}
\end{figure*}
 ListNet uses a single layer feed forward neural network without bias term and nonlinear activation function. We improve the original architecture with recent techniques such as layer normalization \cite{ba2016layer}, highway connections \cite{srivastava2015highway} and exponential linear units \cite{clevert2015fast}. Inspired by the transformer\cite{vaswani2017attention} architecture, we insert a self-attention layer in the middle of two feed forward layers. We will refer to this architecture as document encoder (DE). Figure \ref{fig:docencoder} shows the architecture of document encoder. 

\subsection{ListNet + Regularized Self-Attention (RSA)}
\begin{figure}[htbp!]
    \begin{subfigure}[t]{0.5\linewidth}
        \includegraphics[width=0.9\linewidth]{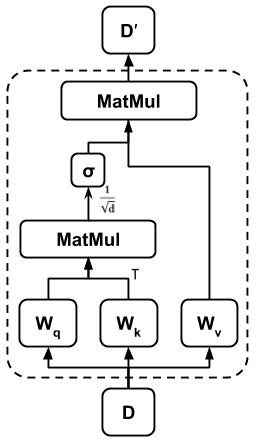}
        \caption{Self-attention}
        \label{fig:self_att}
    \end{subfigure}%
    \begin{subfigure}[t]{0.5\linewidth}
        \includegraphics[width=0.9\linewidth]{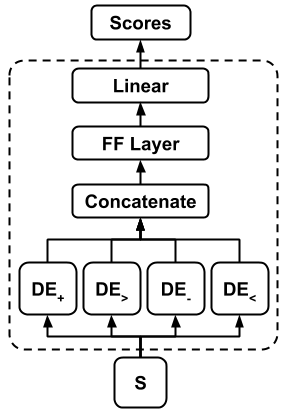}
        \caption{ListNet + RSA}
        \label{fig:sarn}
    \end{subfigure}%
    \caption{Self-attention layer and ListNet + Regularized Self-Attention (RSA).}
\end{figure}
We observe that certain document interactions are embedded in the datasets: 1) relative orderings between documents and 2) arithmetic differences in relevance judgments between documents. We hypothesize that this information can provide powerful supervisions for the learning of the self-attention weights. We explore four different document encoders, each of which is supervised by a different regularization term: 
\begin{itemize}
  \item $\mathbf{DE_+}$ is a document encoder which enhances vector representations of documents by paying attention to other documents that are more relevant. 
  i.e, for a given document $d_i$, the attention weight $W^+_{ij}$ for $d_j$ is:
  \begin{equation}
  W^+_{ij} =
          \begin{dcases}
        1 &\text{if } r_j > r_i \\ \label{eq:de1}
        0 &\text{if } r_j <= r_i \\
    \end{dcases}
  \end{equation}
  \item $\mathbf{DE_>}$ is similar to $DE_+$ except that it assigns exponentially higher attention weights to documents with higher relevance judgments:
\begin{equation}
  W^>_{ij} =
          \begin{dcases}
        \dfrac{e^{(r_j - r_i)}}{\sum_{i=0}^{k}e^i} &\text{if } r_j > r_i \\\label{eq:de2}
        0 &\text{if } r_j <= r_i \\
    \end{dcases}
  \end{equation}
  \item $\mathbf{DE_-}$ does the opposite of $DE_+$. It assigns positive attention weights to documents that are less relevant.
\begin{equation}
  W^-_{ij} =
          \begin{dcases}
        1 &\text{if } r_j < r_i \\\label{eq:de3}
        0 &\text{if } r_j >= r_i \\
    \end{dcases}
  \end{equation}
  \item $\mathbf{DE_<}$ is similar to $DE_-$, except that it assigns exponentially higher attention weights to documents with lower relevance judgments:
\begin{equation}
  W^<_{ij} =
          \begin{dcases}
        \dfrac{e^{(r_i - r_j)}}{\sum_{i=0}^{k}e^i} &\text{if } r_j < r_i \\\label{eq:de4}
        0 &\text{if } r_j >= r_i \\
    \end{dcases}
\end{equation}
\end{itemize}
In equations \eqref{eq:de2} and \eqref{eq:de4}, k refers to the maximum relevance judgment. In this paper, k=4 for all the datasets.

The outputs from the four document encoders are concatenated and then converted to scores via another feedforward layer.
The final scores are used to rank the documents.

\subsection{Regularization Terms}
We introduce regularization terms which encourage the document encoders to learn attention weights close to the values mentioned in equations \eqref{eq:de1}, \eqref{eq:de2}, \eqref{eq:de3} and \eqref{eq:de4}:

Rewrite equation \eqref{eq:1} as:
\begin{equation}
  V' = \Sigma(VW_v)
\end{equation}
where:
\begin{equation}
  \Sigma = \sigma\left((VW_q)(VW_k)^\intercal\right)
\end{equation}
$\Sigma$ is the attention matrix of a document encoder, $\Sigma \in \mathbb{R}^{n \times n}$.\\
The regularization terms are defined as the average binary cross entropy between the attention weight matrices and the ideal attention weight matrices defined in equations \eqref{eq:de1}, \eqref{eq:de2}, \eqref{eq:de3} and \eqref{eq:de4}:
\begin{equation}
\begin{split}
  L_\gamma = -\dfrac{1}{n^2}\sum_{i=1}^n
  &\sum_{j=1}^n 
  [W^\gamma_{ij}log(\Sigma^\gamma_{ij}) 
  + (1 - W^\gamma_{ij})log(1-\Sigma^\gamma_{ij})]\\
\end{split}
\end{equation}
for $\gamma \in \{+, >, -, <\}$.

Final objective function is the summation of the ListNet loss function and the regularization terms:
\begin{equation}
L = L + L_+ +L_> + L_- + L_<
\end{equation}

\section{Experimental Setup}

\subsection{Datasets}

\begin{table*}[!htbp]
  \caption{Characteristics of the datasets. }
  \label{tab:datasetstats} 
  \centering
  \begin{tabular}{|c|c|c|c|c|c|c|}
    \hline
    \textbf{Dataset} & \textbf{Year} &\textbf{\# Features} & \textbf{Type} & \textbf{\#Queries (Q)} & \textbf{\#Documents (D)} & \textbf{Average \# D/Q}\\
    \hline
    \multirow{3}{*}{Yahoo LETOR} & \multirow{3}{*}{2010} & \multirow{3}{*}{700}& Train & 19944 & 473134& 23.72\\
    \cline{4-7}
     &  &  & Validation & 2994  & 71083   & 23.74\\
    \cline{4-7}
    &  &   &Test & 6983   & 165660  & 23.72\\
    \hline
    \multirow{3}{*}{MSLR-WEB10K} & \multirow{3}{*}{2010} & \multirow{3}{*}{136} & Train & 6000 & 723412& 120.57\\
    \cline{4-7}
      &  & &Validation & 2000 & 235259   & 117.63\\
    \cline{4-7}
      &  && Test & 2000   & 241521  & 120.76\\
    \hline
    \multirow{3}{*}{MSLR-WEB30K} & \multirow{3}{*}{2010} & \multirow{3}{*}{136} & Train & 18919 & 2270296& 120.00\\
    \cline{4-7}
      &  & & Validation & 6306    & 747218   & 118.49\\
    \cline{4-7}
      &  & & Test & 6306   & 753611  & 119.51\\
    \hline
    \multirow{3}{*}{Istella-S LETOR} & \multirow{3}{*}{2016} & \multirow{3}{*}{220} & Train & 19245 & 2043304& 106.17\\
    \cline{4-7}
      &  & & Validation & 7211 & 684076   & 118.49\\
    \cline{4-7}
      &  & & Test & 6562   & 681250  & 103.82\\
    \hline
    \multirow{3}{*}{Istella LETOR} & \multirow{3}{*}{2016} & \multirow{3}{*}{220} & Train & 17331 & 5459701 & 315.03\\
    \cline{4-7}
      &  & & Validation & 5888 & 1865924& 316.90\\
    \cline{4-7}
      &  & & Test & 9799 & 3129004& 319.32 \\
    \hline
    \end{tabular}
\end{table*}

We conduct evaluations on the Yahoo LETOR\cite{chapelle2011yahoo}, MSLR-WEB30K \cite{DBLP:journals/corr/QinL13} and Istella LETOR\cite{dato2016fast} datasets shown in table \ref{tab:datasetstats}. 
We also include results on the MSLR-WEB10K and Istella-S LETOR datasets, which are sampled from MSLR-WEB30K and Istella LETOR respectively. 

Due to privacy regulations, all datasets only contain extracted feature vectors and raw texts of queries and documents are not publicly available.
Every dataset has five levels of relevant judgment, from 0 (not relevant) to 4 (highly relevant).

\subsection{Baseline Systems and Parameters Tuning}
All neural models were implemented with PyTorch \footnote{https://pytorch.org/}. 
We also provide results of two strong learning to rank algorithms based on ensembles of regression trees: MART \cite{friedman2002stochastic} and LambdaMART \cite{burges2010ranknet}.
We used RankLib\footnote{https://sourceforge.net/p/lemur/wiki/RankLib/\\We omit the results of other learning to rank algorithms in Ranklib as they perform significantly worse than MART and LambdaMART.} to train and evaluate these models and did hyperparameter tuning on the number of trees and the number of leaves per tree. 

Models with highest NDCG@10 scores on validation sets were used to obtain final results on test sets and significance tests were conducted using paired t-test.

\subsection{Evaluation Metrics}
We consider two popular ranking metrics which support multiple levels of relevance judgment:
\begin{enumerate}
    \item \textbf{Normalized Discounted Cumulative Gain (NDCG) \cite{jarvelin2002cumulated}} sums relevance judgments (gain) of ranked documents, which are discounted by their positions in ranking and normalized by the discounted cumulative gain of the ideal documents ordering. 
    \item \textbf{Expected Reciprocal Rank (ERR) \cite{chapelle2009expected}} measures the expected reciprocal rank at which a user will stop his search.
\end{enumerate}
We report results at positions 1, 3, 5 and 10 for both metrics. 

\section{Results and analyses}
\subsection{Results}
\begin{table*}[!htbp]
\caption{Evaluation results. * and + indicate results which are statistically significant different from the results of ListNet + RSA at p$<$0.01 and 0.01$\leq$p$<$0.05 respectively.}
\medskip
\begin{subtable}{\textwidth}
\centering
  \caption{Results on Yahoo LETOR}
  \label{tab:yahoo}
  \begin{tabular}{|c|c|c|c|c|c|c|c|c|}
    \hline
    \textbf{Algorithm}  & \textbf{ERR@1} & \textbf{NDCG@1} 
                        & \textbf{ERR@3} & \textbf{NDCG@3}  
                        & \textbf{ERR@5} & \textbf{NDCG@5}
                        & \textbf{ERR@10} & \textbf{NDCG@10}\\
    \hline
    MART & 0.3440  & 0.6837 & 0.4196 & \textbf{0.6877} & 0.4400 & \textbf{0.7072} & 0.4553 & 0.7468\\
    \hline
    LambdaMART & \textbf{0.3460} & \textbf{0.6870} & \textbf{0.4205} & 0.6853 & \textbf{0.4409} & 0.7040 & \textbf{0.4553} & 0.7468+\\
    \hhline{|=|=|=|=|=|=|=|=|=|}
    ListNet & 0.3394 & 0.6703 & 0.4140 & 0.6764 & 0.4348 & 0.6974 & 0.4496 & 0.7420* \\
    \hline
    ListNet + SA & 0.3381* & 0.6723* & 0.4127* & 0.6765* & 0.4338* & 0.6967* & 0.4486* & 0.7422* \\
    \hline
    ListNet + RSA & 0.3418 & 0.6733 & 0.4171 & 0.6839 & 0.4382 & 0.7066 & 0.4526 & \textbf{0.7499}\\
    \hline
    \end{tabular}
\end{subtable}

\medskip
\begin{subtable}{\textwidth}
\centering
  \caption{Results on MSLR-WEB10K}
  \label{tab:mslr10k}
  \begin{tabular}{|c|c|c|c|c|c|c|c|c|c|}
    \hline
    \textbf{Algorithm} & \textbf{ERR@1} & \textbf{NDCG@1} 
                        & \textbf{ERR@3} & \textbf{NDCG@3}  
                        & \textbf{ERR@5} & \textbf{NDCG@5}
                        & \textbf{ERR@10} & \textbf{NDCG@10}\\
    \hline
    MART &  0.2172* & 0.4149* & 0.2971* & 0.4161* & 0.3213* & 0.4260* & 0.3399* & 0.4479*\\
    \hline
    LambdaMART & 0.2261* & 0.4278* & 0.3061* & 0.4244* & 0.3291* & 0.4305* & 0.3479* & 0.4510*\\
    \hhline{|=|=|=|=|=|=|=|=|=|}
    ListNet & 0.2108* & 0.4095* & 0.2845* & 0.3978* & 0.3081* & 0.4072* & 0.3276* & 0.4292* \\
    \hline
    ListNet + SA & 0.2127* & 0.4019* & 0.2904* & 0.4050* & 0.3127* & 0.4104* & 0.3315* & 0.4310* \\
    \hline
    ListNet + RSA & \textbf{0.2305} & \textbf{0.4386} & \textbf{0.3099} & \textbf{0.4347} & \textbf{0.3319} & \textbf{0.4383} & \textbf{0.3502} & \textbf{0.4568}\\
    \hline
    \end{tabular}
\end{subtable}

\medskip
\begin{subtable}{\textwidth}
\centering
   \caption{Results on MSLR-WEB30K}
  \label{tab:mslr30k}
  \begin{tabular}{|c|c|c|c|c|c|c|c|c|c|}
    \hline
    \textbf{Algorithm}  & \textbf{ERR@1} & \textbf{NDCG@1} 
                        & \textbf{ERR@3} & \textbf{NDCG@3}  
                        & \textbf{ERR@5} & \textbf{NDCG@5}
                        & \textbf{ERR@10} & \textbf{NDCG@10}\\
    \hline
    MART & 0.2217* & 0.4364+ & 0.3032* & 0.4259 & 0.3276 & 0.4352 & 0.3468* & 0.4574*\\
    \hline
    LambdaMART & 0.2395* & 0.4580+ & 0.3213* & 0.4461 & 0.3442* & 0.4512* & 0.3630* & 0.4711*\\
    \hhline{|=|=|=|=|=|=|=|=|=|}
    ListNet & 0.2288* & 0.4289* & 0.3072* & 0.4241* & 0.3295* & 0.4290* & 0.3481* & 0.4489* \\
    \hline
    ListNet + SA & 0.2270* & 0.4286* & 0.3040* & 0.4215* & 0.3269* & 0.4272* & 0.3460* & 0.4492* \\
    \hline
    ListNet + RSA & \textbf{0.2494} & \textbf{0.4643} & \textbf{0.3264} & \textbf{0.4524} & \textbf{0.3489} & \textbf{0.4568} & \textbf{0.3669} & \textbf{0.4775}\\
    \hline
    \end{tabular}
\end{subtable}

\medskip
\begin{subtable}{\textwidth}
\centering
   \caption{Results on Istella-S LETOR}
  \label{tab:istella_s}
  \begin{tabular}{|c|c|c|c|c|c|c|c|c|c|}
    \hline
    \textbf{Algorithm} & \textbf{ERR@1} & \textbf{NDCG@1} 
                        & \textbf{ERR@3} & \textbf{NDCG@3}  
                        & \textbf{ERR@5} & \textbf{NDCG@5}
                        & \textbf{ERR@10} & \textbf{NDCG@10}\\
    \hline
    MART & 0.5563* & 0.6251* & 0.6739* & 0.6054* & 0.6922* & 0.6334* & 0.6998* & 0.7038*\\
    \hline
    LambdaMART & 0.5868* & 0.6579* & 0.6983* & 0.6309* & 0.7146* & 0.6561* & 0.7213* & 0.7193*\\
    \hhline{|=|=|=|=|=|=|=|=|=|}
    ListNet & 0.5855* & 0.6566* & 0.6981* & 0.6316* & 0.7141* & 0.6577* & 0.7203* & 0.7190* \\
    \hline
    ListNet + SA & 0.5921* & 0.6633* & 0.7025* & 0.6345* & 0.7189* & 0.6618* & 0.7252* & 0.7232* \\
    \hline
    ListNet + RSA & \textbf{0.5986} & \textbf{0.6714} & \textbf{0.7113} & \textbf{0.6490} & \textbf{0.7264} & \textbf{0.6757} & \textbf{0.7322} & \textbf{0.7394}\\
    \hline
    \end{tabular}
\end{subtable}

\medskip
\begin{subtable}{\textwidth}
\centering
   \caption{Results on Istella LETOR}
  \label{tab:istella}
  \begin{tabular}{|c|c|c|c|c|c|c|c|c|c|}
    \hline
    \textbf{Algorithm}  & \textbf{ERR@1} & \textbf{NDCG@1} 
                        & \textbf{ERR@3} & \textbf{NDCG@3}  
                        & \textbf{ERR@5} & \textbf{NDCG@5}
                        & \textbf{ERR@10} & \textbf{NDCG@10}\\
    \hline
    MART & 0.5629* & 0.6208* & 0.6683* & 0.5667* & 0.6864* & 0.5838* & 0.6946* & 0.6323*\\
    \hline
    LambdaMART & 0.5938* & 0.65436* & 0.6962* & 0.5958* & 0.7123* & 0.6104* & 0.7194* & 0.6574* \\
    \hhline{|=|=|=|=|=|=|=|=|=|}
    ListNet & 0.5760* & 0.6327* & 0.6805* & 0.5775* & 0.6964* & 0.5888* & 0.7041* & 0.6317* \\
    \hline
    ListNet + SA & 0.5911* & 0.6499* & 0.6955* & 0.5959* & 0.7108* & 0.6098* & 0.7182* & 0.6556* \\
    \hline
    ListNet + RSA & \textbf{0.6035} & \textbf{0.6646} & \textbf{0.7085} & \textbf{0.6153} & \textbf{0.7236} & \textbf{0.6293} & \textbf{0.7302} & \textbf{0.6777} \\
    \hline
\end{tabular}
\end{subtable}
\end{table*}

Tables \ref{tab:yahoo}, \ref{tab:mslr10k}, \ref{tab:mslr30k}, \ref{tab:istella_s} and \ref{tab:istella} present our main results on the datasets. 
We observe that ListNet with additional self-attention layer only does marginally better than ListNet without self-attention layer.
However, ListNet with regularized self-attention consistently achieves very strong ERR@$i$ and NDCG@$i$ scores at various positions $i$. 
In particular, ListNet with regularized self-attention is the single best system in all metrics measured in four out of the five datasets. For example, in MSLR-WEB10K, our model achieves 0.2305 ERR@1, 0.4386 NDCG@1, 0.3502 ERR@10, and 0.4568 NDCG@10, all outperforming the next-best model LambdaMART (which achieved 0.2261 ERR@1, 0.4278 NDCG@1, 0.3479 ERR@10, 0.4510 NDCG@10). 
This trend holds true for the MSLR-Web30K, Istella-S, and Istella datasets as well. 
The only exception where ListNet + RSA does not win on all metrics is the Yahoo LETOR datasets: but even there ListNet + RSA ranks second or third in most cases, and still outperforms on NDCG@10. 

These consistent improvements confirm that the proposed regularized self-attention mechanism is an effective way to improve learning to rank results.

\subsection{Impact of Regularization Terms}

\begin{figure}[htbp!]
    \centering
    \includegraphics[width=0.99\linewidth]{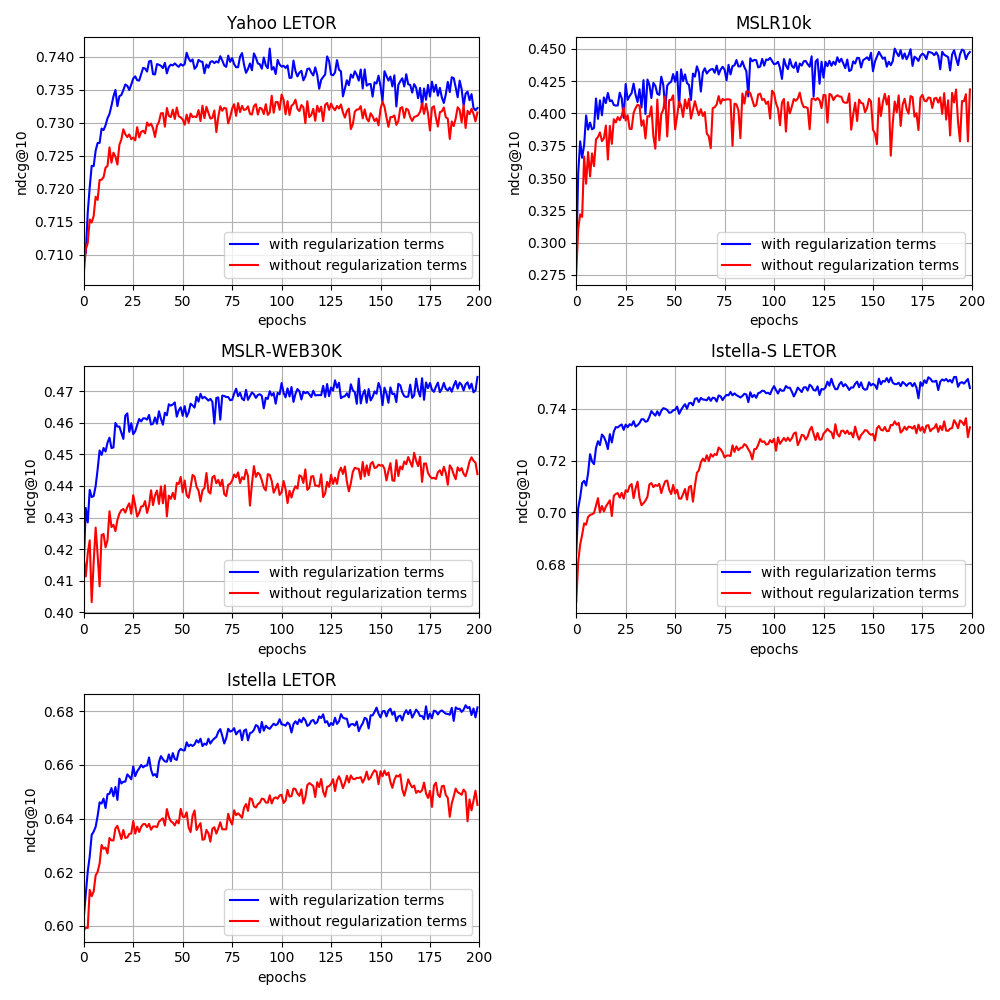}
    \caption{Plots of NDCG@10 scores against training epochs on all validation sets. Curves of models with regularization terms are almost always above the curves of models without regularization terms.}
    \label{fig:no_loss}
\end{figure}
Figure \ref{fig:no_loss} shows the plots of NDCG@10 scores against training epochs on all validation sets. 
As seen in the plots, the curves of models with the regularization terms are almost always above the curves of models without the regularization terms.
Further, the former always converge to significantly higher NDCG@10 values than the latter. 
These phenomenons clearly show that our proposed regularization terms are effective at improving the performance of ListNet with self-attention layer. 
In fact, models without the regularization terms perform worse than MART and LambdaMART on all datasets. 

\subsection{Attention Visualization}
\begin{figure}[htbp!]
    \centering
    \includegraphics[width=\linewidth]{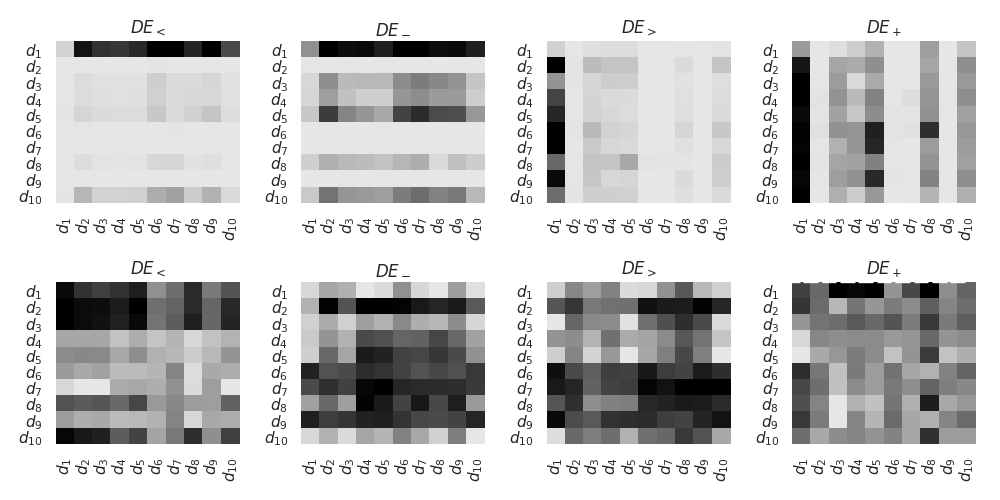}
    \caption{Top row: attention weights matrices. Bottom row: attention weights matrices without regularization terms. The relevance judgments of the documents for this sample query are $d_1=3$, $d_2=0$, $d_3=0$, $d_4=1$, $d_5=3$, $d_6=0$, $d_7=0$, $d_8=1$, $d_9=0$ and $d_{10}=3$.}
    \label{fig:heatmap}
\end{figure}
We sample query and document pairs from the Istella-S dataset and plot the heatmaps of the attention weights of the four different document encoders in figure \ref{fig:heatmap}. 

Bottom row of figure \ref{fig:heatmap} shows attention heatmaps of a model trained without the regularization terms. 
We are unable to observe any explainable pattern in the attention matrices.
From the results our experiments, self-attention alone is not effective at figuring out attention weights that are useful for modeling document interactions.

In contrast, top row of the visualization suggests that our model can learn better attention weights with the supervisions from the regularization terms: $DE_<$ and $DE_-$ place more attention weights on \emph{rows} with higher relevance judgments, while $DE_>$ and $DE_+$ place more attention weights on \emph{columns} with higher relevance judgments. 

\subsection{Impact of the document encoders}
\begin{figure}[htbp!]
    \centering
    \includegraphics[width=0.85\linewidth]{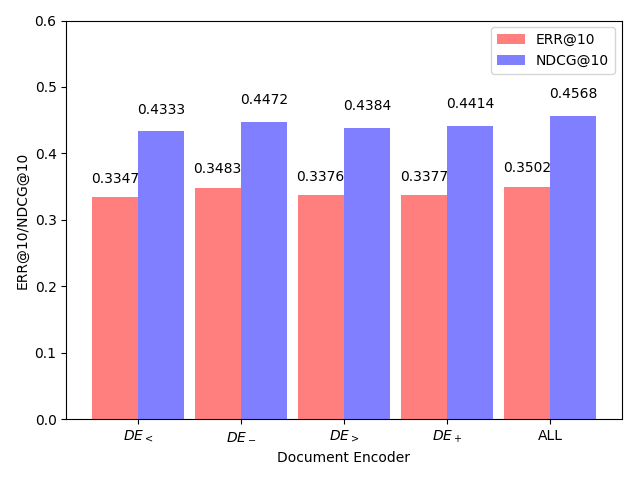}
    \caption{ERR@10 and NDCG@10 scores on the MSLR-WEB10K test set for different document encoders.}
    \label{fig:des}
\end{figure}

We train four separate models, each of which uses only one of the four document encoders. Figure \ref{fig:des} presents our results. 

We observe that $DE_+$ or $DE_-$ perform better than $DE_>$ and $DE_<$, despite the fact that $DE_>$ and $DE_<$ are trained to put exponentially higher attention weights on documents with higher relevance judgments. We suspect the regularization terms for $DE_>$ and $DE_<$ are more difficult to optimize than the regularization terms for $DE_+$ and $DE_-$.

We also observe that the NDCG@10 score from a model with four document encoders is around 1 to 2.3 points higher than the models with individual document encoders. This shows that ensembling the document encoders is effective at improving results on learning to rank tasks.

\section{Related Work}

There are two general research directions in learning to rank. 
In the traditional setting, machine learning algorithms are employed to re-rank documents based on preprocessed feature vectors. 
In the end-to-end setting, models are designed to extract features and rank documents simultaneously.

\subsection{Traditional Learning to Rank}

As there can be tens of millions of candidate documents for every query in real-world contexts, information retrieval systems usually employ a two-phase approach.
In the first phase, a smaller set of candidate documents are shortlisted from the bigger pool of documents using simpler models such as vector space model \cite{salton1975vector} and BM25 \cite{robertson2009probabilistic}. 
In the second phase, shortlisted documents are converted into feature vectors and more accurate learning to rank models are used to re-rank the feature vectors. 
Examples of commonly used features are term frequencies, BM25 scores, URL click rate and length of documents. 

Predicting the relevance scores of documents in pointwise approaches can be treated as a regression problem.
Popular regression algorithms such as \cite{breiman2001random,friedman2002stochastic} are often directly used to estimate relevance judgments of documents. 
Pairwise approaches such as \cite{adomavicius2005toward,joachims2002optimizing, burges2007learning} treat learning to rank as binary classification problem. 
Ensemble trees are generally recognized as the strongest systems, e.g.~an ensemble of LambdaMART and other lambda-gradient models \cite{burges2011learning} won the Yahoo Learning to Rank challenge \cite{chapelle2011yahoo}. 
Neural networks such as RankNet \cite{burges2005learning} and ListNet \cite{cao2007learning} are also effective.
The common theme in these papers is to learn a classifier which can determine the correct ordering given a pair of documents.

Optimizing listwise objectives can be difficult.
This is because popular IR metrics such as MAP, ERR and NDCG are not differentiable, which means gradient descent based methods cannot be directly used for optimization. 
Various surrogate loss functions have been proposed over the years. 
For example, \cite{cao2007learning} proposed ListNet, which uses the cross entropy between the permutation probability distributions of the predicted ranking and the ground truth as loss function. 
\cite{taylor2008softrank} proposed SoftRank, which uses smoothed approximations to ranking metrics. \cite{burges2010ranknet} proposed LambdaRank and LambdaMART, which approximate gradients by the directions of swapping two documents, scaled by change in ranking metrics. 
Although these loss functions demonstrate various degree of success on learning to rank tasks, most of the papers only use them to train global ranking models which predict relevance scores of every document independently. In contrast, our model is designed specifically to model the interdependence between documents. Our ListNet loss function can also be replaced with any of the above existing loss objectives.

More recently, \cite{ai2018learning} proposed the DLCM, which uses recurrent neural network to sequentially encode documents in the order returned by strong baseline learning to rank algorithms such as LambdaMART. The authors find that incorporating local ranking context can further fine-tune the initial results of baseline systems. Unlike DLCM which relies on the ranking results from other learning to rank algorithms, our model is a self-contained learning to rank algorithm. Therefore, direct comparison between DLCM and our model is not possible.

\subsection{End-to-end learning to rank}
As traditional learning to rank systems rely heavily on handcrafted feature engineering that can be tedious and often incomplete, there is growing interest in end-to-end learning to rank tasks among both NLP and IR researchers. 
Systems in this category focus on generating feature vectors automatically using deep neural networks, without the need for feature vectors extraction.

End-to-end models can be further be classified under two broad categories: 1) representation-based models and 2) interaction-based models.
Representation-based models try to generate good representations of query and document independently before conducting relevance matching e.g, \cite{huang2013learning,hu2014convolutional}, while interaction-based models focus on learning local interactions between query text and document text before aggregating the local matching signals, e.g, \cite{guo2016deep,pang2017deeprank,mcdonald2018deep}.

Since the aforementioned models focus primarily on learning better vector representations of query-document pairs from raw texts, the output representations from those models can be directly fed as inputs to our model, which is designed to learn the interactions among the documents. 
As end-to-end learning to rank is not the focus of this paper, we will explore end-to-end models in future work.

\section{Conclusions}
This paper explores the possibility of modeling document interactions with self-attention mechanism. We show that a self-attentional neural network with properly regularized attention weights can outperform state-of-the-art learning to rank algorithms on publicly available benchmark datasets.

We believe that different interpolation weights in the regularization terms of the loss function in Equation 18 may affect performance; we will explore this in future work. 
Another line of future work is to combine the idea of self-attention proposed here to the various end-to-end deep learning ranking models proposed in the literature.

\bibliographystyle{ACM-Reference-Format}
\bibliography{bib}
\end{document}